\documentclass{moriond}

\def\Journal#1#2#3#4{{#1} {\bf #2}, #3 (#4)}

\def\EPJC{{\em Eur. Phys. J.} C}

\def\be{\begin{equation}}
\def\ee{\end{equation}}
\def\bea{\begin{eqnarray}}
\def\eea{\end{eqnarray}}

\newcommand{\Photo}{\includegraphics[height=35mm]{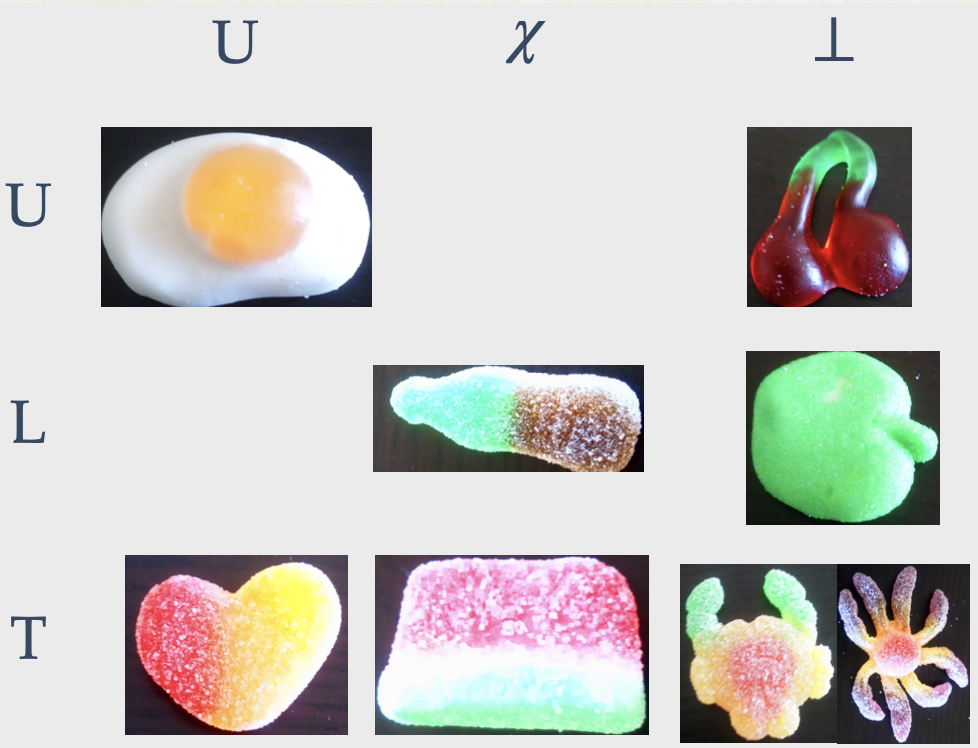}}

\begin{document}
\vspace*{4cm}
\title{The vector boson transverse momentum distributions}

\author{Ignazio Scimemi}

\address{Departamento de Física Teórica and IPARCOS, Facultad de Ciencias Físicas, Universidad Complutense Madrid, Plaza Ciencias 1, 28040 Madrid, Spain}

\maketitle\abstracts{
The transverse momentum dependent distributions (TMD) are an essential part of the factorization theorems in  vector boson production. They are non-perturbative, double scale dependent functions that asymptotically match onto collinear parton distributions functions (PDF). Once TMD are expressed using PDF, one observes  that they are sensitive to the choice and quality of PDF sets ({\it{PDF bias}}). A solution to this problem is found and discussed. Nevertheless the main source of error on vector boson spectra still comes from PDF uncertainty propagation. }
The motion of quarks and gluons inside a hadron affects the transverse momentum dependent cross sections in Drell-Yan (DY), semi-inclusive deep inelastic scattering (SIDIS) and semi-inclusive annihilation (SIA). The factorized differential cross sections of these processes, $d\sigma$, reveals that the partons inside a hadron organize themselves into spin (in)dependent distributions called TMD, $f$ \footnote{ In the talk I associated each TMD to a jelly candy, which can be extracted from a bag (the hadron) and distinguished one from the other using an appropriate observable. This analogy motivates the front picture.}.
For the unpolarized processes when vector boson transverse momentum, $q_T$,  is fixed and $q_T \ll Q$, with $Q$ the hard energy scale, one finds~\cite{Collins:2011zzd,Echevarria:2011epo}
\begin{equation} 
\label{partonfact} 
{{d \sigma} \over { dQ^2 dyd q_T^2} } =  \sum_{i,j}  \int d^2 b \ e^{i \bf{b  \cdot q_T}} \sigma^{(0)}_{i j} f_{1,i\leftarrow h1}(x_1,b;\mu,\zeta_1)f_{1,j\leftarrow h2}(x_2,b;\mu,\zeta_2) , 
\end{equation} 
with $q$ the vector boson momentum, $b=\sqrt{\bf{b}^2}\geq 0$  the transverse distance Fourier conjugate to  $q_T$, $Q^2\sim q^2$ and $ \sigma^{(0)}_{i j}$ the perturbatively calculable partonic cross sections. The TMD $f_{i,j}$  depend on the mass and rapidity scales $\mu$ and $\zeta$ that obey the corresponding evolution equations, and 
they are, by definition, non-perturbative. All scales dependence can be collected into factors using the so called $\zeta$-prescription~\cite{Scimemi:2018xaf},
\begin{equation} 
\label{zetafact} 
{{d \sigma} \over { dQ^2 dyd q_T^2} } =  \sum_{i,j}  \int d^2 \mathbf{b} \ e^{i \bf{ b  \cdot q_T}} \sigma^{(0)} H_{i j}(Q,Q) R(b;Q,  Q^2) f_{1,i\leftarrow h1}(x_1,b)f_{1,j\leftarrow h2}(x_2,b), 
\end{equation}
where  $H_{i,j}(Q,\mu=Q)$ is the perturbatively calculable hard factor, $R(b;\mu=Q,\sqrt{\zeta_1\zeta_2}=Q^2)$ the evolution kernel, and $f_{i,j}$ the scale independent part of TMD. The $\zeta$-prescription is the only one which can achieve this type of factorization formula and it also ensures a high convergence of the perturbative series. The evolution kernel $R$ is valid universally for DY, SIDIS and SIA and it is flavor independent.  All flavor dependence can be collected inside the TMD,
\begin{eqnarray}\label{eq:fnp}
f_{1,f\leftarrow h}(x,b)=\sum_{f'} f^{\rm NP}_{f\leftarrow h}(x,b)\int_x^1\frac{dy}{y} C_{f\leftarrow f'}(y,\mathbf{L}_{\mu_{\rm OPE}},a_s(\mu_{\rm OPE}))\; f_{f\leftarrow h}(x/y,\mu_{\rm OPE})\;
\end{eqnarray}
being $C_{f\leftarrow f'}$ a perturbatively calculable Wilson coefficient and $f_{f\leftarrow h}$ a collinear PDF. In eq.~(\ref{eq:fnp})  the limit 
$$\lim_{b\rightarrow 0}f_{1,f\leftarrow h}(x,b)=\sum_{f'} \int_x^1\frac{dy}{y} C_{f\leftarrow f'}(y,\mathbf{L}_{\mu_{\rm OPE}},a_s(\mu_{\rm OPE}))\; f_{f\leftarrow h}(x/y,\mu_{\rm OPE})\;$$
holds. The function $f^{\rm NP}_{f\leftarrow h}(x,b)$ collects non-perturbative flavor dependent contributions beyond the PDF ones and so it represents the original ingredient of TMD with respect to  standard resummation. Because $f^{\rm NP}_{f\leftarrow h}(x,b)$  depends on two variables one has to use a large  number of experimental results (at very different values of $Q$) to extract it on a relevant portion of the $(x,b)$ plane. While several groups are taking on this challenge, mine has observed that the flavor dependence of $f^{\rm NP}_{f\leftarrow h}(x,b)$ is essential to achieve an agreement among different PDF sets~\cite{Bury:2022czx}. In other words if one neglects the flavor dependence of  $f^{\rm NP}_{f\leftarrow h}$ one finds  a strong PDF set sensitivity of the $\chi^2$ of the  fit. This issue is actually a {\it PDF bias}, as the collinear non-perturbative physics issues can affect strongly the extractions of multidimensional  distributions.

A major part of the study performed in ref.~\cite{Bury:2022czx} concerns understanding the source and amount of errors in the TMD extractions, using the public code {\it artemide}~\cite{web,Scimemi:2017etj} which includes all theoretical necessary inputs.
The statistically relevant errors come from experimental data and PDF replicas, while scale errors indicate the expected contributions from higher order perturbative calculations. 
 The experimental uncertainty is evaluated  by generating pseudo-data~\cite{Ball:2008by}. The replicas of the pseudo-data are obtained adding Gaussian noise to the values of data points and scaling uncertainties if required. The noise parameters are driven by experimental correlated and uncorrelated uncertainties.
 The uncertainty due to the collinear PDF is accounted for by using each PDF replica as input and  including 1000 replicas.
In our previous studies~\cite{Bertone:2019nxa,Scimemi:2019cmh,Hautmann:2020cyp}  it has been concluded that the NNLO perturbative inputs~\cite{Gehrmann:2014yya,Echevarria:2015usa,Echevarria:2015byo,Echevarria:2016scs,Luo:2019hmp} account reasonably for the theoretical part, and not much improvement is expected  by the 3-loops contributions that have been recently  calculated~\cite{Luo:2019szz,Luo:2020epw,Ebert:2020yqt,Vladimirov:2016dll,Li:2016ctv}. An update of the code including these new inputs is nevertheless in progress.  The perturbative inputs considered here for TMD  are at next-to-next leading order (NNLO), (i.e. all coefficients and anomalous dimensions at order $\alpha_s^2$ and the cusp anomalous dimension at order $\alpha_s^3$)  and considered  PDF sets are CT18~\cite{Hou:2019efy}, HERA20~\cite{H1:2015ubc}, MSHT20~\cite{Bailey:2020ooq}, NNPDF31~\cite{NNPDF:2017mvq}.
The PDF values and  their evolution 
are taken from the LHAPDF \cite{Buckley:2014ana}. In order to speed up the analysis a first study is performed  on a reduced set of DY data, which are the most sensitive to TMD and finally  the complete data sets are included, confirming the results. 

The PDF and experimental errors are evaluated with the method of replicas on different PDF sets. The evidence of the PDF bias has appeared
re-examining the SV19 extraction~\cite{Scimemi:2019cmh}. There, firstly,
for each PDF set it is evaluated the error generated by its 1000 replicas~\cite{NNPDF:2017mvq} keeping $f^{\rm{NP}}$ fixed. Then  fitting $f^{\rm{NP}}$ for each PDF replica, the same conclusion is reached: the inclusion of the PDF uncertainty produces a broader, more realistic band for the TMDPDF.
It has been also interesting to observe that in each set,
most of the PDF replicas (more than $75\%$) have $\chi^2/N_{\rm{pt}}>2$, while the central replica describes the data with $\chi^2/N_{\rm{pt}}\sim 1$ and this occurs whether or not we fit $f_{\rm{NP}}$ for each replica. The issue is common to all set of data. The main consequence of this fact is that a reweighing of replicas cannot be an solution for improving our analysis.

\begin{figure}
\begin{center}
\begin{minipage}{0.45\linewidth}
\centerline{\includegraphics[width=0.7\linewidth]{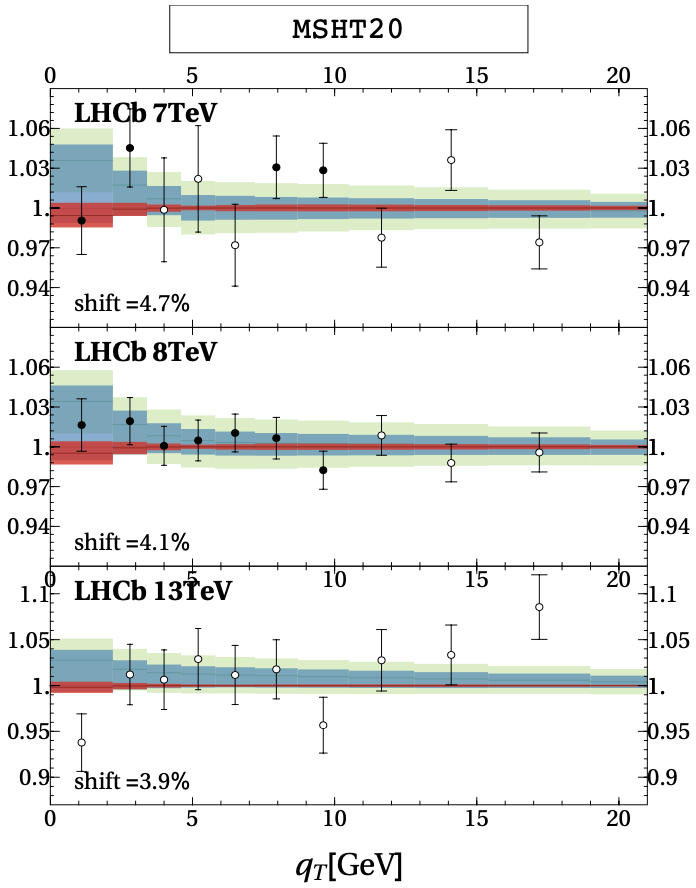}}
\end{minipage}
\begin{minipage}{0.45\linewidth}
\centerline{\includegraphics[width=0.7\linewidth]{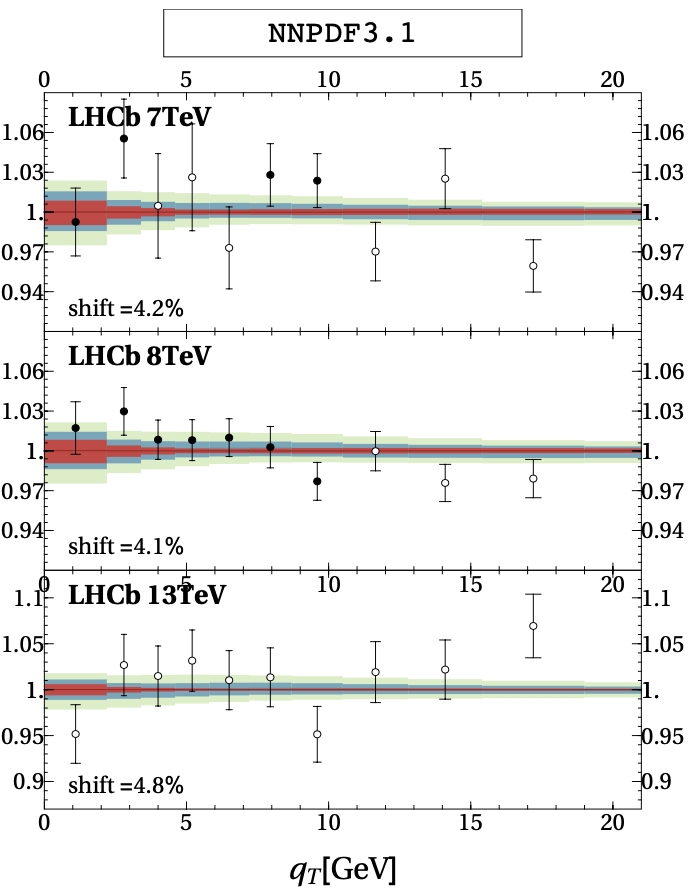}}
\end{minipage}
\hfill
\caption[]{Example of the data description at high energy. Left panel: the ratio $d\sigma_{\rm{experiment}}/d\sigma_{\rm{theory}}$ for Z-boson production  by the LHCb~\cite{LHCb:2016fbk,LHCb:2015okr,LHCb:2015mad} experiment with two different sets of PDF.  The red band is the \textbf{EXP}-uncertainty. The light-green band is the \textbf{PDF}-uncertainty. The blue band is the combined uncertainty. Only the filled bullets are included into the fit.
More plots on Drell-Yan experimental results can be found in ref.~\cite{Bury:2022czx}.
}
\label{fig:lhcb}
\end{center}
\end{figure}

The observed PDF bias is highly reduced introducing a flavor dependent ansatz for the $f_{\rm{NP}}$.  In ref.~\cite{Bury:2022czx}
this consists of
\begin{eqnarray}\label{def:fNP}
f_{\rm{NP}}^{f}=\exp\left(-\frac{(1-x)\lambda_1^f+x\lambda_2^f}{\sqrt{1+\lambda_0 x^2 b^2}}b^2\right),
\end{eqnarray}
with $\lambda_{1,2}^f>0$ and $\lambda_0>0$. 
The model has  a Gaussian shape at intermediate $b$ followed by an exponential asymptotic fall at $b\to\infty$.
The factor $\lambda_0$ accompanying $x^2$ is common to previous SV19 extraction (that found  ${x=2.05\pm 0.25}$) and follows the  general pattern of $b$ power corrections to TMDPDF $\sim (xb)^2$, suggested in ref.~\cite{Moos:2020wvd}. 
In order to keep a low number of parameters the parameters $\lambda_{1,2}$ are taken  to be flavor dependent while $\lambda_0$ is universal for all flavors. 
 The  $u$, $d$, $\bar u$, $\bar d$ and $sea$ cases are distinguished among each other and the  $sea$ made of $(s, \bar s, c, \bar c, b ,\bar b)$  flavors obtaining a total of 11 free parameters. The final $\chi^2$ for each PDF set are
1.12 (MSHT20), 0.91 (HERA20), 1.21 (NNPDF31), 1.08 CT18. The spread of $\chi^2$ value among replicas is also highly reduced. Nevertheless the PDF error is still actually the main source of uncertainty. As an example we show the LHCb case in fig.~\ref{fig:lhcb} taken from the supplementary material of the original paper~\cite{Bury:2022czx}, where more details are also provided.
 In tab.~\ref{tab:parameters} the fitted parameters show a substantial agreement among different sets of PDF, within uncertainties.
 We have checked that the recent results of fiducial distributions at CMS~\cite{CMS:2021oex} (see also the talk of B. Bilin in the electro-weak session of Moriond 2022) and LHCb~\cite{LHCb:2021huf} at $s=\sqrt{13 }$ TeV are correctly predicted using the settings of this work (and will be shown in future publications).
 
\begin{table}[t]
\begin{center}
\renewcommand{\arraystretch}{1.1}
\begin{tabular}{||c  |c | c | c | c ||}
\hline
Parameter & \textbf{MSHT20}  & \textbf{HERA20}  & \textbf{NNPDF31}  & \textbf{CT18}  \\
\hline \hline
$\lambda_1^u$ & $0.12_{-0.04}^{+0.12}$ & $0.11_{-0.07}^{+0.07}$ & $0.28_{-0.10}^{+0.12}$ & $0.05_{-0.05}^{+0.09}$
\\
$\lambda_2^u$ & $0.32_{-0.22}^{+1.84}$ & $8.15_{-3.51}^{+2.09}$ & $2.58_{-2.05}^{+1.37}$ & $0.9_{-0.71}^{+0.84}$
\\\hline
$\lambda_1^d$ & $0.37_{-0.10}^{+0.09}$ & $0.44_{-0.31}^{+0.09}$ & $0.40_{-0.22}^{+0.10}$ & $0.29_{-0.22}^{+0.11}$
\\
$\lambda_2^d$ & $1.7_{-1.6}^{+2.4}$ & $0.11_{-0.11}^{+1.14}$ & $1.1_{-1.0}^{+2.4}$ & $4.7_{-4.4}^{+5.0}$
\\\hline
$\lambda_1^{\bar u}
\times 100$ & $0.37_{-0.27}^{+2.51}$ & $11.6_{-7.6}^{+7.1}$ & $8.8_{-8.6}^{+10.1}$ & $0.94_{-0.85}^{+8.14}$
\\
$\lambda_2^{\bar u}$ & $56._{-12.}^{+6.}$ & $6.5_{-6.4}^{+5.5}$ & $13._{-6.}^{+28.}$ & $56._{-22.}^{+4.}$
\\\hline
$\lambda_1^{\bar d} \times 100$ 
& $0.12_{-0.02}^{+0.18}$ & $35._{-15.}^{+6.} $ & $9.8_{-9.5}^{+9.4}$ & $0.12_{-0.02}^{+0.86}$
\\
$\lambda_2^{\bar d}$ & $1.1_{-1.0}^{+2.4}$ & $0.05_{-0.05}^{+0.09}$ & $6.1_{-1.6}^{+16.7}$ & $0.37_{-0.26}^{+2.20}$
\\\hline
$\lambda_1^{s}$ & $0.11_{-0.10}^{+0.07}$ & $0.49_{-0.15}^{+0.45}$ & $0.25_{-0.12}^{+0.15}$ & $0.012_{-0.011}^{+0.006}$
\\
$\lambda_2^{s}$ & $5.1_{-3.4}^{+3.3}$ & $5.2_{-5.0}^{+7.7}$ & $3.0_{-3.0}^{+4.3}$ & $9.1_{-4.5}^{+4.2}$
\\\hline
$\lambda_0$ & $29._{-10.}^{+53.}$ & $339._{-212.}^{+156.}$ & $181._{-133.}^{+39.}$ & $24_{-11.}^{+73.}$
\\\hline
$c_0\times 100$ & $4.36_{-0.31}^{+0.31}$ & $3.27_{-0.52}^{+1.23}$ & $2.61_{-0.61}^{+0.97}$ & $4.39_{-0.44}^{+0.41}$
\\\hline
\end{tabular}
\caption{\label{tab:parameters} The values of the NP parameters obtained in the fit.}
\end{center}
\end{table}

\section*{Acknowledgments}
I would like to acknowledge all the support from my collaborators F. Hautmann, S. Leal Gomez, A. Vladimirov, P. Zurita.
The work has been possible thanks to the Spanish Ministry grant PID2019-106080GB-C21 and European Union Horizon 2020 research and innovation program under grant agreement Num. 824093 (STRONG-2020).

\end{document}